\documentclass[conference]{IEEEtran}
\IEEEoverridecommandlockouts

\usepackage{cite}
\usepackage{amsmath,amssymb,amsfonts}
\usepackage{algorithmic}
\usepackage{algorithm}
\usepackage{graphicx}
\usepackage{textcomp}
\usepackage{xcolor}
\usepackage{booktabs}
\usepackage{array}
\usepackage[hyphens]{url}
\usepackage{listings}

\def\BibTeX{{\rm B\kern-.05em{\sc i\kern-.025em b}\kern-.08em
    T\kern-.1667em\lower.7ex\hbox{E}\kern-.125emX}}

\newcommand{\sysname}{LLM-Emu}

\begin{document}

\pdfpagewidth=8.5in
\pdfpageheight=11in

\newcommand{\iscasubmissionnumber}{7}

\pagenumbering{arabic}

\title{\sysname{}: Native Runtime Emulation of LLM Inference via Profile-Driven Sampling}
\author{\IEEEauthorblockN{Wei Da}
\IEEEauthorblockA{\textit{University of Cambridge}\\
Cambridgeshire, United Kingdom \\
wd312@cam.ac.uk}
\and
\IEEEauthorblockN{Evangelia Kalyvianaki}
\IEEEauthorblockA{\textit{University of Cambridge}\\
Cambridgeshire, United Kingdom \\
ek264@cam.ac.uk}}

\maketitle
\thispagestyle{plain}
\pagestyle{plain}

\begin{abstract}
Realistic evaluation of LLM serving systems requires online workloads, dynamic arrivals, queueing, and the serving engine’s local scheduling for execution batching, but running such experiments on GPUs is expensive. Existing simulators reduce this cost, but often operate offline or in time-warped mode, re-implement serving-engine schedulers, or require accurate operator/kernel-level latency models. We present \sysname{}, a serving-native emulator for vLLM that preserves the production HTTP, scheduling, KV-cache, and output-processing paths while replacing only GPU forward execution with profile-sampled latency and synthetic output tokens. Tested on two different GPUs, four model variants, two model families, two attention backends, and both Poisson and bursty ShareGPT workloads, \sysname{} closely tracks real vLLM serving behavior: TPOT and ITL stay within $4.8\%$ absolute error, E2E latency within $5.3\%$, and output throughput within $1.9\%$; TTFT is less stable, with maximum error $10.4\%$, reflecting its sensitivity to admission and queue state. These results suggest that lightweight, serving-native emulation can support practical online experimentation for LLM-serving systems. \sysname{} is open sourced at \textcolor{blue}{\url{https://github.com/AKafakA/llm-emu}}.
\end{abstract}

\section{Introduction}

Large language models (LLMs) have demonstrated remarkable capabilities across a wide range of tasks and applications, driving demand for efficient serving systems and clusters in both industry and academia~\cite{kwon2023vllm,yu2022orca,agrawal2024sarathi, zhang2023flashinfer, zhong2024distserve,sun2024llumnix}. However, developing and evaluating improvements on serving engines across deployment scales and dynamic online environments usually requires expensive hardware resources. 


Existing simulators and emulators reduce the cost of studying LLM serving systems, but they still leave gaps for realistic online experimentation ~\cite{agrawal2026revati, agrawal2024vidur, feng2025frontier, cho2025llmservingsim}. Many systems are offline simulators or accelerated configuration-search tools rather than wall-clock serving endpoints, so they cannot directly exercise live HTTP traffic, dynamic arrivals, queueing behavior, and runtime overheads in the deployed serving stack ~\cite{agrawal2024vidur,apex,nvidia2026aiconfigurator,cho2025llmservingsim,feng2025frontier}.  Their performance models also introduce several assumptions. For example, some rely on analytical or operator-level abstractions ~\cite{nvidia2026aiconfigurator,apex,cho2025llmservingsim,feng2025frontier}, while others use learned predictors over profiled operators ~\cite{agrawal2024vidur,agrawal2026revati}. These designs are effective for fast design-space exploration, but can be difficult to explain, calibrate, and generalize across workloads, hardware, and serving-engine versions. In addition, several simulators re-implement or model serving-engine behavior against particular engine assumptions or versions ~\cite{agrawal2024vidur,cho2025llmservingsim}. As serving engines evolve, these parallel implementations must be manually kept in sync. Reproducibility is also uneven as some of recent systems do not provide a public implementation, making it difficult to inspect assumptions, reproduce results, or leverage them to test new settings
~\cite{agrawal2026revati,feng2025frontier}. Recent emulation work such as Revati~\cite{agrawal2026revati} executes real serving-framework code, but targets accelerated time-warped emulation through CUDA interception and predicted kernel durations rather than wall-clock online serving.

\begin{figure}[t]
  \centering
  \includegraphics[width=\columnwidth]{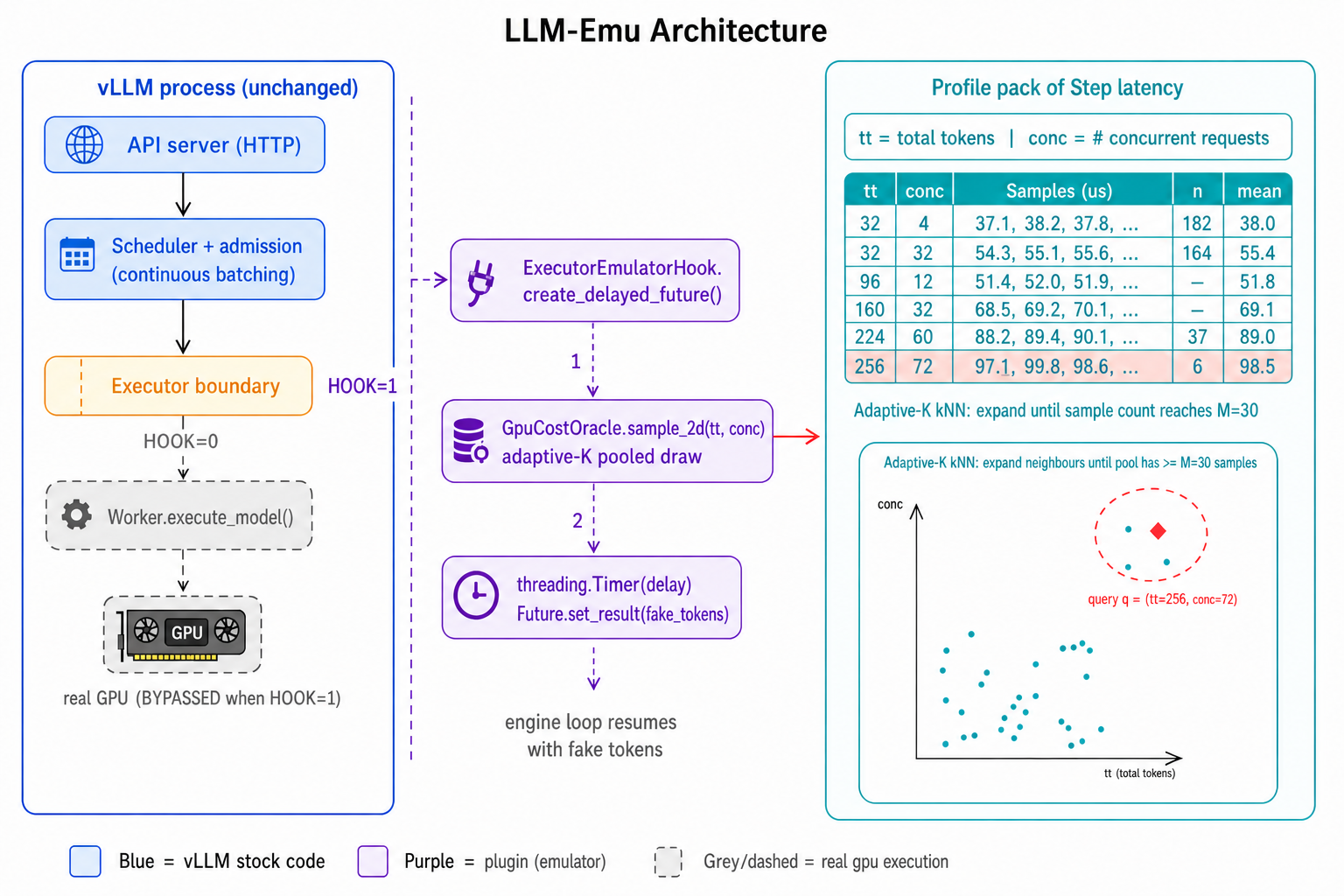}
  \caption{\sysname{} plugs in at the executor boundary; everything
  else is vLLM's own code.}
  \label{fig:arch}
\end{figure}

To address these gaps, we present \sysname{}, a profile-driven online emulator that runs inside a stock vLLM process as a wall-clock runtime plugin. \sysname{} preserves vLLM's scheduler, HTTP stack, admission path, KV-cache management, and output pipeline, and replaces only the GPU forward path with a sampled latency from offline profiles keyed by batch shape and concurrency. This design avoids a parallel scheduler implementation, per-operator latency modeling, and CUDA interception, while allowing the same vLLM server CLI and HTTP clients to run against either real or emulated execution.

\sysname{} implements this idea with a density-aware profile oracle and a timer-resolved \texttt{Future} that returns synthetic output tokens after the predicted delay. We evaluate \sysname{} across two GPUs, four model variants from two model families, two attention backends, default vLLM configurations, and ShareGPT online serving workloads under multiple request rates and arrival patterns. Across these settings, \sysname{} closely tracks real vLLM behavior for TPOT, ITL, E2E latency, and throughput, with larger but explainable TTFT error due to queueing and startup sensitivity.

This paper makes three contributions. First, we introduce a serving-native emulation design that preserves the production vLLM online serving path and replaces only GPU forward execution. Second, we design a lightweight density-aware latency oracle and timer-resolved Future to preserve asynchronous scheduler-worker overlap. Third, we validate \sysname{} across hardware, model scale, model family, attention backend, and arrival distribution, showing low error for steady-state serving metrics.

\section{Background and Related work}
\label{sec:background}

\subsection{LLM Serving Anatomy.}
Large language models (LLMs) are typically Transformer-based autoregressive models. In serving, inference is commonly divided into two phases: \emph{prefill}, which processes the input prompt and materializes the corresponding key/value tensors into the KV cache, and \emph{decode}, which generates tokens iteratively while reusing cached keys and values from prior tokens. Serving engines batch multiple requests to improve GPU utilization, but request arrivals, completions, sequence lengths, and KV-cache growth make the batch shape and memory footprint change at every iteration.

Modern serving systems address this dynamism with iteration-level scheduling and continuous batching, as introduced by Orca~\cite{yu2022orca} and widely adopted in systems such as vLLM~\cite{kwon2023vllm}. vLLM further uses PagedAttention to manage KV cache in fixed-size blocks through a page-table-like mapping, reducing fragmentation and enabling flexible allocation, preemption, recomputation, or swapping under memory pressure. Chunked prefill further splits long prefills into smaller chunks that can be interleaved with decode steps to reduce latency and improve utilization. In practice, the serving framework sits between the HTTP/API layer and GPU kernel libraries such as FlashAttention~\cite{dao2022flashattention} and FlashInfer~\cite{zhang2023flashinfer}, handling admission, scheduling, memory management, and output processing.

The inference serving system is usually used for runtime applications with dynamic requests and arrivals called the HTTP client and API. To measure the online performance, several metrics are commonly used. Time to first token (TTFT) measures the latency from receiving a request to generating the first token. Time per output token (TPOT) measures the average time to generate each output token after the first one. Iteration time latency (ITL) measures the time taken for each iteration of the generation process. End-to-end latency (E2E) measures the total time from receiving a request to generating the complete response. Output-token throughput (TPS, token per second) measures the number of generated tokens per unit time.

\subsection{Existing simulators and emulators.}
Due to the rapid growth of LLM applications, LLM serving systems have become an active area of research and development. However, developing and evaluating LLM serving systems requires GPU resources, leading to high cost and long iteration cycles. To address this issue, several simulators and emulators have been proposed to reduce dependence on real hardware for early evaluation.

Vidur~\cite{agrawal2024vidur} mirrors the scheduling logic of an inference serving system to simulate batch composition and request concurrency at each step. It then trains a random forest model to predict per-batch latency from operator-level features, enabling end-to-end workload-trace simulation across cluster SKU configurations. AIConfigurator~\cite{nvidia2026aiconfigurator} targets fast configuration search for LLM serving deployments by combining analytical performance models, calibrated kernel-level data, and backend-aware configuration generation, rather than executing the production serving engine. APEX~\cite{apex} is a CPU-side simulator for selecting parallel execution plans; it abstracts model execution, batching, quantization, and device clusters while modeling basic iteration-level batching behavior. LLMServingSim~\cite{cho2025llmservingsim} is a system-level simulator for heterogeneous LLM serving infrastructure, using trace-driven performance modeling and operator-level latency profiling while exposing simulator-side interfaces for routing, cache management, and scheduling policies. Frontier~\cite{feng2025frontier} further extends this simulator family toward MoE expert parallelism and prefill/decode/AF disaggregation.

These systems are useful for capacity planning and design-space exploration, but they share three practical limitations. First, they either re-implement serving-engine behavior or omit parts of it, making them vulnerable to drift as engines such as vLLM evolve. Second, they rely on learned, analytical, or learned latency models that require calibration and can be difficult to generalize across workloads, hardware, and engine versions. Third, most operate as offline simulators, so they cannot directly exercise live HTTP traffic, dynamic arrivals, queueing behavior, and deployed-stack overheads in a wall-clock online serving setting. This also creates a reproducibility and maintenance gap. To our knowledge at the time of writing, publicly available simulator-style tools do not execute the current vLLM V1 online serving path with recent features such as chunked prefill and asynchronous scheduling. Vidur is open source, but its public implementation follows an earlier vLLM-V0 engine path; LLMServingSim and APEX similarly require simulator-side support for new engine features; and Frontier's implementation is not publicly available.

Revati~\cite{agrawal2026revati} takes a different approach. Instead of reimplementing the serving scheduler, it executes real serving-framework code while intercepting CUDA API calls through \texttt{LD\_PRELOAD}. Rather than running GPU kernels, Revati advances a virtual clock by predicted kernel durations and synchronizes these time jumps across distributed workers, reporting 5--17$\times$ faster-than-real-time emulation with less than 5\% prediction error. This design preserves much of the serving-engine control logic, but it targets offline time-warped emulation rather than a wall-clock online endpoint. As a result, it is less suited for directly exercising online overheads and dynamics. Revati also depends on CUDA interposition and kernel-duration prediction, which may require calibration and ongoing maintenance as CUDA libraries, hardware platforms, workloads, and serving-engine implementations evolve. In addition, no public artifact is available to reproduce or extend its results at the time of writing.

\sysname{} is motivated by this gap. Rather than re-implementing the serving engine or intercepting CUDA calls, \sysname{} keeps vLLM's deployed online serving stack on the real code path and replaces only GPU forward execution with profile-sampled latency and synthetic output tokens. This design targets a different point in the design space which includes wall-clock online emulation with minimal integration surface, direct compatibility with existing vLLM online serving, and reduced maintenance burden as vLLM evolves.

\section{\sysname{}}
\label{sec:design}

\subsection{Overview}

\begin{figure}[t]
  \centering
  \includegraphics[width=\columnwidth]{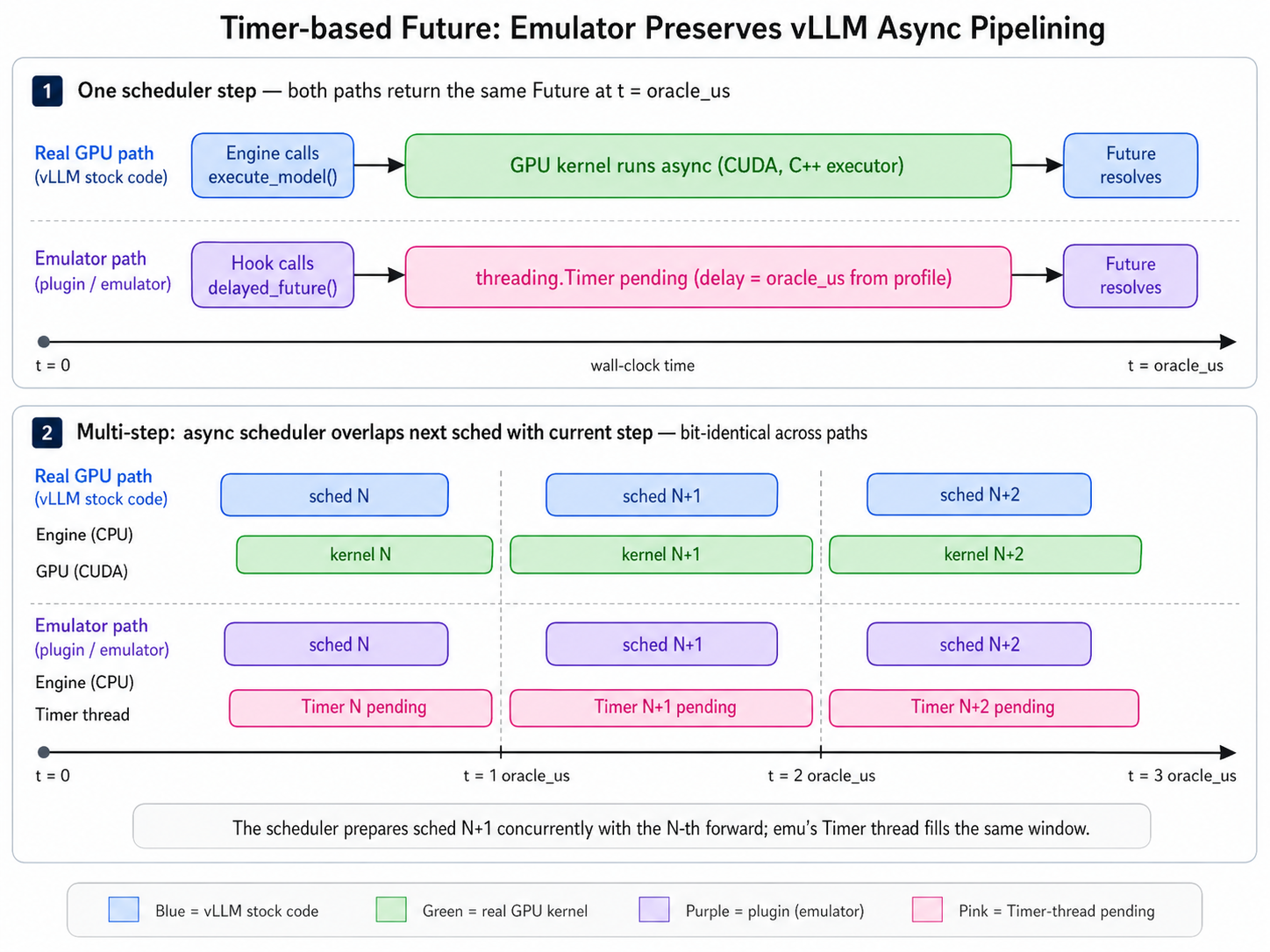}
  \caption{Timer-based Future preserves scheduler--worker overlap.}
  \label{fig:timeline}
\end{figure}

Figure~\ref{fig:arch} shows the \sysname{} architecture and its integration with vLLM for emulation. \sysname{} is implemented as a runtime plugin that hooks into the GPU worker's per-step execution path. When enabled, the plugin redirects the model runner through an emulated execution path consisting of a density-aware latency oracle and a timer-resolved \texttt{Future}. The oracle samples a latency from a two-dimensional profile indexed by the current batch's total token count and request concurrency; sparse regions are filled in by an adaptive nearest-neighbor expansion (Algorithm~\ref{alg:adaptive-k}). \sysname{} then schedules the \texttt{Future} to resolve after the predicted latency and returns synthetic output token IDs to the model runner for downstream post-processing. This asynchronous path is used for the online vLLM server configuration targeted by this work, as shown in Figure~\ref{fig:timeline}. For non-asynchronous execution, such as offline batch inference through the \texttt{LLM()} interface, \sysname{} falls back to a blocking wait path. We implement this mode for completeness but do not evaluate it in this paper.

This design emulates GPU execution time while leaving the rest of vLLM unchanged. As a result, request admission, scheduling, KV-cache management, output processing, and HTTP-serving logic remain on the production vLLM code path and behave as they would in a real deployment. The serving-native, sampling-based design also avoids the per-operator latency tables and learned-predictor calibration required by Vidur, AIConfigurator, and LLMServingSim~(\S\ref{sec:background}). Supporting a new model, hardware platform, or configuration only requires collecting a new profile, as described in \S\ref{sec:design-profile}.

\subsection{Offline Profiling}
\label{sec:design-profile}

The profile pack used by the oracle is a JSON artifact capturing per-step latency as two joint distributions (decode-only and prefill-or-mixed) over two-dimensional buckets keyed by \textit{tt} (total tokens in the step) and \textit{conc} (concurrency, the number of running requests). A third combined step-cycle table is retained as a sparse-bucket fallback. We choose total tokens and concurrency to capture the dominant variation in per-step GPU execution time under continuous batching: aggregate token work and batch shape. We further split decode-only and mixed prefill/decode steps because these phases have different latency behavior. This intentionally simple feature set trades generality for low calibration cost and interpretability. Each bucket stores the raw list of observed latencies rather than a pre-aggregated summary, so that the oracle can resample over per-sample neighbors at query time as used by the density-aware Shepard pooling in Algorithm~\ref{alg:adaptive-k}, and preserve real variance.

For the offline profile capture, we run a sweep across request rates from light load to saturation, with denser sampling at higher rates where batch composition is most volatile. A representative profile used for the main experiments is $\approx$5.9\,MB of JSON containing $\approx$276K samples across $\approx$7.3K $(tt, \text{conc})$ buckets, captured in 3.5 to 4.5 hours of GPU wall-clock time across two seeded rounds of the rate sweep. We use the same dataset and the same vLLM CLI flags for both profile capture and evaluation; profile capture may issue more prompts per rate than evaluation, but the workload shape and flag set are identical, so the profile's sample distribution matches the workload the oracle is asked to predict at runtime. An alternative is a synthetic sweep that explicitly enumerates $(tt, \text{conc})$ buckets independent of the workload's distribution, but it is far more expensive and left for future work.

\begin{algorithm}[t]
\caption{Density-aware neighbor pooling}
\label{alg:adaptive-k}
\begin{algorithmic}[1]
\STATE \textbf{input:} query $(t,c)$, bucket set $B$, reliability floor $M$
\STATE sort $B$ by range-normalized 2D distance to $(t,c)$
\STATE $\mathcal{S} \leftarrow \emptyset$,\;\; $n \leftarrow 0$
\FOR{$B_i$ in sorted order}
  \STATE $\mathcal{S} \leftarrow \mathcal{S} \cup \{B_i\}$
  \STATE $n \leftarrow n + |B_i.\text{samples}|$
  \IF{$n \ge M$} \STATE \textbf{break} \ENDIF
\ENDFOR
\RETURN Shepard-weighted sample over $\mathcal{S}$
\end{algorithmic}
\end{algorithm}

\subsection{Implementation}
\label{sec:design-impl}

\sysname{} targets vLLM~0.18.1 and is implemented as a small runtime plugin. A lazy-import hook in \path{vllm/v1/worker/gpu_worker.py} installs the emulated execution path during GPU-worker initialization. When the plugin is enabled, the hook bypasses model loading and real GPU setup, allowing the vLLM server to start in a GPU-free emulation mode. It then routes each executor step to the \texttt{vllm\_emulator/} package, which contains the profile loader, the density-aware oracle, and the timer-resolved \texttt{Future} return path. The hook is controlled by environment variables that enable the oracle and select the profile pack, so emulation can be enabled alongside the normal vLLM server CLI. Switching between real and emulated serving is therefore a one-line launch-time change, and downstream HTTP clients, including \texttt{vllm bench serve}, require no modification:

\begin{lstlisting}[basicstyle=\ttfamily\scriptsize,breaklines=true,columns=fullflexible]
# Emulated serve, same CLI
$ VLLM_EMULATOR_ENABLE_ORACLE=1 \
  VLLM_EMULATOR_PROFILE_PACK=profile.json \
# Same as Real vLLM serve
  vllm serve Qwen/Qwen3-8B \
    --max-model-len 4096 --port 8100
\end{lstlisting}

\sysname{} keeps the runtime footprint small. The total delta on top of vLLM 0.18.1 is ${\approx}2.5$K lines as 1.7K lines of online emulator (oracle, executor hook, GPU-free startup shims and platform plugin), and 600 lines of offline profiling tools (per-step tracer plus profile-pack builder), and 173 LoC wiring inside existing vLLM codebase that activates the plugin and routes per-step traces. For relative scale, Vidur~\cite{agrawal2024vidur} is ${\approx}11$K Python, LLMServingSim (2.0 version)~\cite{cho2025llmservingsim} is ${\approx}15$K lines, AIConfigurator~\cite{nvidia2026aiconfigurator} is ${\approx}83$K lines including embedded hardware-latency tables, and Revati~\cite{agrawal2026revati} includes ${\approx}6.9$K lines of C++ for CUDA interception plus per-framework patches. Such a qualitative comparison of LoC enables us to assess that \sysname{}'s code paths avoid scheduler re-implementation, per-operator latency modeling, and CUDA-interception machinery.

\section{Evaluation}
\label{sec:results}

\begin{table}[t]
\centering
\caption{Per-cell Relative error $(\text{emu}-\text{real})/\text{real}$ across six cells and five request rates. Bold entries mark the maximum absolute error for each metric.}
\label{tab:main}
\small
\begin{tabular}{@{}lrrrrr@{}}
\toprule
Metric & $r{=}2$ & $r{=}4$ & $r{=}8$ & $r{=}16$ & $r{=}32$ \\
\midrule
\multicolumn{6}{@{}l}{\textbf{M-Q8 --- Main: Qwen3-8B at RTX~8000}} \\
TTFT & $-9.81\%$ & $-6.92\%$ & $-1.89\%$ & $+3.17\%$ & $+2.52\%$ \\
TPOT & $+0.48\%$ & $+0.85\%$ & $+0.18\%$ & $+0.08\%$ & $+0.80\%$ \\
ITL  & $+0.43\%$ & $+0.80\%$ & $-0.00\%$ & $-0.03\%$ & $+0.71\%$ \\
E2E  & $+0.34\%$ & $+0.72\%$ & $-0.42\%$ & $+2.01\%$ & $+2.02\%$ \\
TPS & $-0.00\%$ & $-0.02\%$ & $+0.67\%$ & $+0.12\%$ & $-0.78\%$ \\
\midrule
\multicolumn{6}{@{}l}{\textbf{M-Q14 --- Model-scale up: Qwen3-14B at RTX~8000}} \\
TTFT & $-9.10\%$ & $-4.42\%$ & $+1.39\%$ & $+1.24\%$ & $+1.80\%$ \\
TPOT & $+0.47\%$ & $+1.37\%$ & $-1.02\%$ & $+0.27\%$ & $+0.58\%$ \\
ITL  & $+0.42\%$ & $+1.28\%$ & $-1.33\%$ & $+0.39\%$ & $+0.36\%$ \\
E2E  & $+1.06\%$ & $+1.50\%$ & $+0.48\%$ & $+1.09\%$ & $+1.51\%$ \\
TPS & $+0.54\%$ & $-0.26\%$ & $+1.10\%$ & $-0.27\%$ & $-0.02\%$ \\
\midrule
\multicolumn{6}{@{}l}{\textbf{A40-Q8 --- Main (Hardware swap): Qwen3-8B at A40}} \\
TTFT & $-7.37\%$ & $-9.22\%$ & $-1.06\%$ & $+5.05\%$ & $+3.44\%$ \\
TPOT & $+0.94\%$ & $+1.05\%$ & $-0.46\%$ & $+1.24\%$ & $+1.52\%$ \\
ITL  & $+0.82\%$ & $+1.03\%$ & $-0.62\%$ & $+1.26\%$ & $+1.25\%$ \\
E2E  & $+0.74\%$ & $+0.92\%$ & $-0.73\%$ & $+3.72\%$ & $+2.83\%$ \\
TPS & $-0.02\%$ & $+0.02\%$ & $+1.25\%$ & $-1.10\%$ & $-1.03\%$ \\
\midrule
\multicolumn{6}{@{}l}{\textbf{M-Q8-Burst --- Bursty workload: $\gamma{=}0.25$}} \\
TTFT & $-6.90\%$ & $+4.97\%$ & \textbf{-10.41\%} & $-3.13\%$ & $-2.45\%$ \\
TPOT & $+0.08\%$ & \textbf{+4.75\%} & $-2.39\%$ & $-0.90\%$ & $-1.60\%$ \\
ITL  & $-0.14\%$ & \textbf{+4.69\%} & $-2.34\%$ & $-0.93\%$ & $-1.26\%$ \\
E2E  & $-0.20\%$ & $+4.77\%$ & $-4.09\%$ & $-2.32\%$ & $-2.12\%$ \\
TPS & $-0.47\%$ & $-0.71\%$ & \textbf{-1.88\%} & $+0.04\%$ & $+1.01\%$ \\
\midrule
\multicolumn{6}{@{}l}{\textbf{A40-L8 --- Model-family swap: Llama-3.1-8B at A40 }} \\
TTFT & $-6.52\%$ & $-9.32\%$ & $-2.30\%$ & $+4.38\%$ & $+2.12\%$ \\
TPOT & $+0.65\%$ & $+0.46\%$ & $+0.30\%$ & $+1.98\%$ & $+0.73\%$ \\
ITL  & $+0.57\%$ & $+0.38\%$ & $+0.36\%$ & $+1.92\%$ & $+0.61\%$ \\
E2E  & $+0.50\%$ & $+0.28\%$ & $-0.09\%$ & $+3.48\%$ & $+1.69\%$ \\
TPS & $-0.01\%$ & $-0.04\%$ & $-0.75\%$ & $-1.76\%$ & $-0.45\%$ \\
\midrule
\multicolumn{6}{@{}l}{\textbf{A40-Q4 --- Model-scale down: Qwen3-4B at A40}} \\
TTFT & $-6.88\%$ & $-5.16\%$ & $-3.10\%$ & $+7.78\%$ & $+4.80\%$ \\
TPOT & $+0.34\%$ & $+1.28\%$ & $+3.10\%$ & $+1.70\%$ & $+1.13\%$ \\
ITL  & $+0.24\%$ & $+1.21\%$ & $+3.01\%$ & $+1.68\%$ & $+0.79\%$ \\
E2E  & $+0.16\%$ & $+1.14\%$ & $+3.02\%$ & \textbf{+5.22\%} & $+3.60\%$ \\
TPS & $-0.00\%$ & $+0.01\%$ & $-0.23\%$ & $-1.50\%$ & $-0.66\%$ \\
\bottomrule
\end{tabular}
\end{table}

\subsection{Setup}

We evaluate \sysname{} against real vLLM execution on two 48\,GB GPUs: an RTX~8000 using the FlashInfer backend and an A40 using the FlashAttention~2 backend. The RTX~8000 host uses an Intel Xeon Gold 5218 CPU (8~vCPUs at 2.3\,GHz), 62\,GiB RAM, PCIe Gen~3 $\times$16 (8\,GT/s), Ubuntu~24.04, and CUDA driver~590.48. The A40 host is a Vast.ai VM with an AMD EPYC-class 64C/128T CPU, 251\,GiB RAM, PCIe Gen~4 $\times$16 (16\,GT/s), containerized Ubuntu~24.04, and CUDA driver~570.86.

All experiments use stock vLLM~0.18.1 settings for both profile capture and evaluation, including the default online serving path with prefix caching, chunked prefill, and asynchronous scheduling enabled when selected by vLLM. We use the official \texttt{vllm bench serve} ShareGPT workload with 2000 prompts per request rate. Each real run is compared against an emulated run with the same prompts, seed, and request rate, using Poisson arrivals at $r\in\{2,4,8,16,32\}$ unless otherwise stated.

To avoid confounding from run-to-run differences in KV-cache capacity, we record the number of available GPU blocks during profiling and pass the same value to both the real-GPU run and the emulated run via \texttt{--num-gpu-blocks-override}. This keeps memory pressure and preemption behavior aligned between real and emulated serving. We implement this safeguard because, for large models where the KV cache is the primary memory component (e.g., Qwen3-14B), the available block count fluctuated across different boots of the same experiment. Such fluctuations affected the measured deltas, ultimately compromising emulation accuracy.

We evaluate six cells. \textbf{M-Q8} is the main cell, and the remaining five vary one axis at a time:

\begin{itemize}
\item \textbf{M-Q8} (main): Qwen3-8B on RTX~8000 with FlashInfer;
\item \textbf{M-Q14} (model-scale up): Qwen3-14B on RTX~8000;
\item \textbf{M-Q8-Burst} (workload shape): Qwen3-8B on RTX~8000 with bursty arrivals via \texttt{--burstiness 0.25} ($\gamma{=}0.25$ gamma-distributed inter-arrival times). A smaller $\gamma$ produces higher inter-arrival variance and therefore more bursty traffic.
\item \textbf{A40-Q8} (hardware swap): Qwen3-8B on A40 with FlashAttention~2;
\item \textbf{A40-Q4} (model-scale down): Qwen3-4B on A40;
\item \textbf{A40-L8} (model-family swap): Llama-3.1-8B on A40.

\end{itemize}

For the Llama cell, we pass \texttt{--ignore-eos} because first-turn ShareGPT prompts often trigger natural EOS much earlier than the benchmark's reference output length, which is used as the generation cap.

\subsection{Accuracy}
\label{sec:results-main}

Table~\ref{tab:main} reports the per-cell relative error
$(\text{emu}-\text{real})/\text{real}$ for each metric at each rate, across the six cells and five rates.

Across all six cells and five rates, \sysname{} closely tracks real vLLM behavior for steady-state serving metrics. TPOT and ITL stay within 4.8\% absolute error, E2E latency within 5.3\%, and output throughput within 1.9\%. TTFT is less stable, with a maximum absolute error of 10.41\%, because it is more sensitive to admission timing, queue state, and startup effects such as CUDA graph capture or cache warmup.

The largest non-TTFT errors appear under bursty arrivals with $\gamma{=}0.25$, where rapid changes in queue depth create more volatile batch composition than the static profile can fully capture. Even in this setting, TPOT, ITL, E2E latency, and throughput remain within roughly $5\%$, suggesting that the sampling oracle captures the dominant GPU execution behavior across model scale, hardware, workload shape, and model-family changes.

\section{Limitations and future work}
\label{sec:limitations}

\sysname{} provides a high-accuracy, low-maintenance emulator, but our current validation is limited to single-node deployments and matched profile/evaluation workloads. In addition, \sysname{} still requires several hours of GPU time to collect a profile for each model, hardware platform, and serving configuration. We plan to extend \sysname{} in several directions. (a) Reducing profile cost by checking the relationship between different axes to reduce redundancy; (b)~implementing a time-warped accelerated path in the style of Revati~\cite{agrawal2026revati}; (c)~conducting multi-GPU and multi-node validation and extending \sysname{} into a cluster-level emulator for large-scale deployments with target cluster configuration to empower capacity planning and load balancing research. And also (d)~testing with simulation of offline inference by \texttt{LLM()} interface and measuring throughout prediction errors. (e)~Automating API drift detection and oracle revalidation across vLLM releases to reduce manual debugging when vLLM APIs change. Although the lightweight plugin design already minimizes the API surface exposed to drift, automated validation would make this process more robust.

\section{Conclusion}
\label{sec:conclusion}

\sysname{} shows that an online, serving-native LLM emulator can be both simple and accurate. Implemented as a vLLM plugin, \sysname{} replaces only the GPU forward path with a density-aware 2D latency oracle and a timer-resolved Future, while leaving other components on the production vLLM code path. Across variations in model scale, hardware, workload shape, model family, and request load, \sysname{}closely tracks real vLLM serving behavior with only a small executor-boundary hook, rather than broad source modifications. These results suggest that lightweight, serving-native emulation is a promising path toward cheaper and more reproducible online LLM-serving experimentation.

\bibliographystyle{IEEEtranS}
\bibliography{refs}

\end{document}